\documentclass{jetpl}

\usepackage{amssymb}
\usepackage{amsmath}
\usepackage{graphicx}

\DeclareMathOperator{\sgn}{sgn}

\twocolumn \lat

\begin{document}

\title{Josephson effect in SIFS\ tunnel junctions with domain walls in weak
link region }
\rtitle{Josephson effect in SIFS\ tunnel junctions with domain walls in weak
link region}
\sodtitle{Josephson effect in SIFS\ tunnel junctions with domain walls in weak
link region}

\author{S.\,V.~Bakurskiy$^{\,a, \,b}$,
A.\,A~Golubov$^{\,c, \,b}$,
N.\,V.~Klenov$^{\,a}$,
M.\,Yu.~Kupriyanov\/\thanks{mkupr@pn.sinp.msu.ru}$^{\,d, \,b, \,f}$,
I.\,I.~Soloviev$^{\,d}$}

\rauthor{S.\,V.~Bakurskiy, A.\,A~Golubov, N.\,V.~Klenov, et al.}
\sodauthor{Bakurskiy, Golubov, Klenov, et al.}

\address{
$^{a}$ Physics Department, Lomonosov Moscow State University,
 Leninskie gory,  Moscow 119991, Russia\\
 $^{b}$ {Moscow Institute of Physics and Technology, State
 University, Dolgoprudniy, Moscow region, Russia}\\
$^{c}${Faculty of Science and Technology and MESA+, Institute
for Nanotechnology, University of Twente,  Enschede, The
Netherlands}\\
$^{d}$ Skobeltsyn Institute of Nuclear Physics, Moscow, Lomonosov Moscow State University,
 Leninskie gory,  Moscow 119991, Russia\\
$^f$ Institute of Physics, Kazan (Volga region) Federal University, Kremlevskaya ul. 18, Kazan, 420008 Russia}

\date{\today }

\abstract{

We study theoretically the properties of SIFS type Josephson junctions composed of two superconducting (S) electrodes separated by an
insulating layer (I) and a ferromagnetic (F) film consisting of periodic  magnetic domains structure with
antiparallel magnetization directions in  neighboring domains.
The two-dimensional problem in the weak link area is solved analytically in the framework of the linearized quasiclassical Usadel equations.
Based on this solution, the spatial distributions of the critical current
density, $J_{C},$ in the domains and critical current, $I_{C},$ of SIFS structures are calculated as a function of domain wall parameters, as well as the thickness, $d_{F},$ and the width, $W,$ of the domains.
We demonstrate that $I_{C}(d_{F},W)$ dependencies exhibit damped oscillations with the ratio of the decay length, $\xi_{1},$ and oscillation period, $\xi_{2},$ being
a function of the parameters of the domains, and this ratio may take any value from zero to unity. Thus, we propose a new physical mechanism that
may explain the essential difference between $\xi_{1}$ and $\xi_{2}$ observed experimentally in various types of SFS Josephson junctions.
}


\maketitle


It is well known that properties of Josephson structures with ferromagnetic
(F) material in a weak link region depends on relation between the complex
decay length, $\xi,$
( $\xi^{-1} =\xi _{1}^{-1}+i\xi _{2}^{-1})$
and geometrical parameters of these
junctions \cite{RevG}-\cite{RevV}. If F metal is in the dirty limit and exchange energy, $H,$
sufficiently exceeds the critical temperature of superconducting (S)
electrodes, $\pi T_{C},$ then from Usadel equations it follows that $\xi
_{1}\approx \xi _{2}.$ However, it was  demonstrated experimentally \cite{Kontos}-\cite{Blum} that
there could be a noticeable difference between $\xi _{1}\ $and $\xi _{2}.$
Previously the difference has been attributed either to the presence of
strong paramagnetic scattering in the F layer \cite{ryazanov1}, or to violation of the dirty
limit conditions in ferromagnetic material \cite{Blum}, \cite{Pugach}. However, application of the first of
the mechanisms for the experimental data interpretation requires  the existence
of unreasonably strong paramagnetic scattering in the weak link material \cite{ryazanov1}.
The relation between an electron mean free, $\ell ,$ and $%
\xi _{1},$ $\xi _{2}$ in typical experimental situation is also closer to the
dirty limit conditions, $\ell \lesssim \xi _{1},$ $\xi _{2}$ rather than to the clean one.

In this article we prove that the existence of a ferromagnetic domain walls in
F layer can also lead to appearance of substantial differences between $\xi _{1}\ $ and $%
\xi _{2}$ even in the absence of strong scattering by paramagnetic
impurities, and under the fulfilment of the dirty limit conditions in the F material.
\begin{figure}[h]
\centerline{\includegraphics[width=0.5\textwidth]{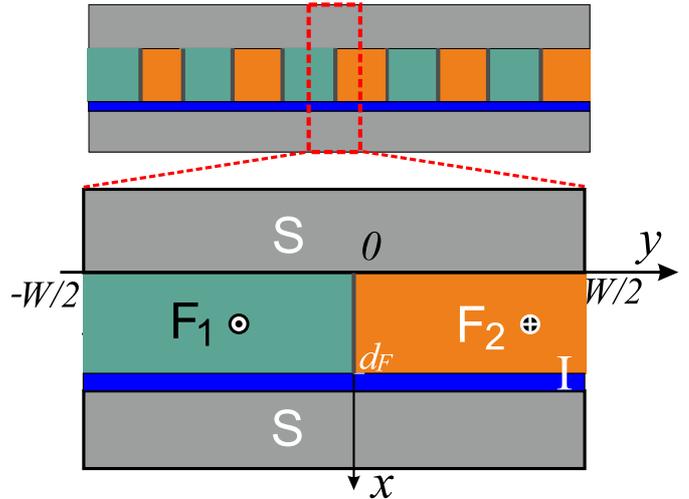}}
\caption{Fig.\protect\ref{fig:fig1}. Geometry of the considered SIFS Josephson junction
and its enlarged part, which includes  two halves of domains and domain wall separating them.
The  insulating barrier I has a small transparency  (shown by a blue
line).  }
\label{fig:fig1}
\end{figure}

\textbf{Model.}
Consider multilayered SIFS structure presented in
Fig.1.
It consists of
superconductor electrode (S), insulator (I) and FS bilayer as an upper
electrode. We assume that the F film has a
thickness, $d_{F},$ and that it subdivides into  domain structure with antiparallel direction of magnetization vector in the neighboring domains.
The width of the domains is $W$ and they separated by atomically sharp domain walls oriented perpendicular to SF interfaces.
Due to periodicity of the structures we, without any loss of generality, can perform our analysis within its half of the period,  
that is from $-W/2$ to $W/2.$ This element is enlarged in Fig.1.
It consists of two halves of domains and domain wall separating them.

We will suppose that the condition of dirty limit is
fulfilled for all metals and that effective electron-phonon coupling
constant is zero in F material.
We will assume further that either temperature $T$ is close to the critical
temperature of superconducting electrodes $T_{C}$ or the suppression parameters $\gamma
_{BS}=R_{BS}\mathcal{A}_{BN}/\rho _{F}\xi _{F} $ at SF interface is large enough to permit the use of the linearized Usadel
equations in F film of the structure. We will characterize  the FF interface (domain wall)  by the suppression
parameter $\gamma=1,$ and the suppression parameter $\gamma _{BF}=R_{BF}\mathcal{A}_{BF}/\rho _{F}\xi _{F},$
which can take any value.
Here $R_{BS},R_{BF}$ and $\mathcal{A}_{BN},\mathcal{A}_{BF}$ are the
resistances and areas of the SF and FF interfaces, $\xi _{S},$ and $\xi
_{F}=(D_{F}/2\pi T_{C})^{1/2}$ are the decay lengths of S, F materials,
while $\rho _{S}$  and $\rho _{F}$ are their resistivities, $%
D_{F}$ is diffusion coefficient in the F metal.

Under the above conditions the
proximity problem in the SF part of SIFS junction $(0\leq x\leq d_{F})$ reduces to solution of the
set of linearized Usadel equations  \cite{RevG}-\cite{RevV}, \cite{Usadel}%
\begin{eqnarray}
\left\{ \frac{\partial ^{2}}{\partial x^{2}}+\frac{\partial ^{2}}{\partial
y^{2}}\right\} F_{F}-\widetilde{\Omega }_{+}F_{F} &=&0,~0\leq y\leq \frac{W}{%
2},  \label{EqFfp1} \\
\left\{ \frac{\partial ^{2}}{\partial x^{2}}+\frac{\partial ^{2}}{\partial
y^{2}}\right\} F_{F}-\widetilde{\Omega }_{-}F_{F} &=&0,~-\frac{W}{2}\leq
y\leq 0,  \label{EqFfp2}
\end{eqnarray}%
where $\Omega =\omega /\pi T_{C},\widetilde{\Omega }_{\pm }=|\Omega |\pm ih%
\sgn(\omega )$, $h=H/\pi T_{C},$  $H,$ is exchange energy of
ferromagnetic material, $\omega =\pi T(2n+1)$ are Matsubara frequencies. The
spatial coordinates in (\ref{EqFfp1}), (\ref{EqFfp2}) are normalized on
decay length $\xi _{F}$. To write these equations  we have chosen
the $x$ and $y$ axis in the directions perpendicular and parallel to the SF
plane and put the origin in the middle of SF interface to the point,
which belongs to the domain wall (see Fig.1).

Equations (\ref{EqFfp1}), (\ref{EqFfp2}) must be supplemented by the
boundary conditions \cite{KL}. They have the form
\begin{eqnarray}
\gamma _{BS}\frac{\partial }{\partial x}F_{F} &=&-G_{0}\frac{\Delta }{\omega
},\ x=0,~-\frac{W}{2}\leq y\leq \frac{W}{2},  \notag \\
\frac{\partial }{\partial x}F_{F} &=&0,\ x=d_{F},~-\frac{W}{2}\leq y\leq
\frac{W}{2}.  \label{BCN(0)}
\end{eqnarray}%
At FF interface $(y=0,~0\leq x\leq d_{F})$ and in the middle of the domains $(y=\pm W/2, ~0\leq x\leq d_{F})$
we also have
\begin{equation}
\gamma _{BF}\frac{\partial }{\partial y}F_{F}(x,+0)=F_{F}(x,+0)-F_{F}(x,-0),%
\label{BCF(0)}
\end{equation}%
\begin{equation*}
\frac{\partial }{\partial y}F_{F}(x,+0)=\frac{\partial }{\partial y}%
F_{F}(x,-0),
\end{equation*}%
\begin{equation}
\frac{\partial }{\partial y}F_{F}(x,\frac{W}{2})=\frac{\partial }{\partial y}%
F_{F}(x,-\frac{W}{2})=0.  \label{BCW}
\end{equation}%
Here $W$ is the width of the domains, $G_{0}=\omega /\sqrt{\omega
^{2}+\Delta ^{2}},$ $\Delta $ is the modulus of the order parameter of
superconducting electrodes. The critical current density, $J_{C},$ of SIFS Josephson junction is
determined by s-wave superconducting correlations at IF interface, which is
even function of the Matsubara frequencies%
\begin{equation}
\frac{eJ_{C}R_{N}}{2\pi T_{C}}=\frac{T}{WT_{C}}\sum_{\omega >0}\frac{%
G_{0}\Delta }{\omega }\Phi(y),  \label{currentD}
\end{equation}%
where $\Phi(y)= (F_{F,+\omega }(d_{F},y)+F_{F,-\omega
}(d_{F},y))/2,$ while the full critical current, $I_{C},$ is the result of integration of $J_{C}(y)$ over
width of the junction.%
\begin{equation}
\frac{eI_{C}R_{N}}{2\pi T_{C}}=\frac{T}{WT_{C}}\sum_{\omega >0}\frac{%
G_{0}\Delta }{\omega }\int_{-W/2}^{W/2}\Phi(y) dy.  \label{FullC}
\end{equation}%
Here, $R_{N},$ is the normal junction resistance.

\textbf{Solution of Usadel equations in FS electrode.}
Solution of two-dimensional boundary value problem (\ref{EqFfp1})-(\ref{BCW}%
) in the F layer $(0\leq x\leq d_{F})$ is convenient to find in the form of the Fourier series expansion%
\begin{equation}
F_{F}(x,y)=\sum_{n=-\infty }^{\infty }A_{n}(y)\cos \frac{\pi nx}{d_{F}}%
,~0\leq y\leq \frac{W}{2},  \label{Fpl}
\end{equation}%
\begin{equation}
F_{F}=\sum_{n=-\infty }^{\infty }B_{n}(y)\cos \frac{\pi nx}{d_{F}},~-\frac{W%
}{2}\leq y\leq 0,  \label{Fmi}
\end{equation}%
where
\begin{eqnarray}
A_{n}(y) &=&\frac{Z}{q_{+}^{2}}+a_{n}\cosh (q_{+}\left( y-\frac{W}{2}\right)
),  \label{A} \\
B_{n}(y) &=&\frac{Z}{q_{-}^{2}}+b_{n}\cosh (q_{-}\left( y+\frac{W}{2}\right)
),  \label{B}
\end{eqnarray}%
and coefficients $a_{n}$ and $b_{n}$
\begin{equation}
a_{n}=-\left[ \frac{1}{q_{+}^{2}}-\frac{1}{q_{-}^{2}}\right] \frac{%
Zq_{-}S_{-}}{\delta%
}, ~q_{\pm }=\sqrt{\widetilde{\Omega }_{\pm }+\left( \frac{\pi n}{d_{F}}\right) ^{2}}, \label{an}
\end{equation}%
\begin{equation}
b_{n}=\left[ \frac{1}{q_{+}^{2}}-\frac{1}{q_{-}^{2}}\right] \frac{Zq_{+}S_{+}%
}{\delta},~ Z=\frac{\Delta G_{0}}{\gamma _{BS}d_{F}\omega }
\label{bn}
\end{equation}%
are determined from boundary conditions (\ref{BCF(0)}). Here the coefficients $\delta,$
$C_{\pm }$ and $S_{\pm }$ are defined by expressions
\begin{equation}
\delta=q_{-}q_{+}\gamma _{BF}S_{+}S_{-}+q_{-}C_{+}S_{-}+q_{+}S_{+}C_{-},
\end{equation}%
\begin{equation}
C_{\pm }=\cosh (\frac{q_{\pm }W}{2}),~S_{\pm }=\sinh (\frac{q_{\pm }W}{2}%
).%
\label{koeff}
\end{equation}%
Taking into account the symmetry relation $q_{-}(-\omega )=q_{+}(\omega )$
for  s-wave superconducting component in the F layer at $x=d_{F}$ it is easy to get
\begin{equation}
\Phi(y\geq0)=\frac{Z}{2}\sum_{n=-\infty }^{\infty }(-1)^{n}\left[ \frac{1}{q_{+}^{2}}+%
\frac{1}{q_{-}^{2}}-
\left[ \frac{1}{q_{+}^{2}}-\frac{1}{q_{-}^{2}}\right]
\frac{\delta_{+}}{\delta}\right],   \label{splus}
\end{equation}%
\begin{equation}
\Phi(y\leq0)=\frac{Z}{2}\sum_{n=-\infty }^{\infty }(-1)^{n}\left[ \frac{1}{q_{+}^{2}}+%
\frac{1}{q_{-}^{2}}-\left[ \frac{1}{q_{+}^{2}}-\frac{1}{q_{-}^{2}}\right]
\frac{\delta_{-}}{\delta}\right],  \label{sminus}
\end{equation}%
\begin{equation*}
\delta _{\pm }=q_{-}S_{-}\cosh (q_{+}\frac{2y\mp W}{2})-q_{+}S_{+}\cosh
(q_{-}\frac{2y\mp W}{2}).
\end{equation*}

Finally for the critical current from (\ref{FullC}), (\ref{splus}) and (\ref%
{sminus}) we have%
\begin{equation}
\frac{eI_{C}R_{N}}{2\pi T_{C}}=\frac{T}{2WT_{C}}\sum_{\omega >0}\frac{%
ZG_{0}\Delta }{\omega }S(\omega ),\label{CurrF}
\end{equation}%
\begin{equation*}
S(\omega )=\sum_{n=-\infty }^{\infty }(-1)^{n}\left[ \frac{W}{q_{+}^{2}}+%
\frac{W}{q_{-}^{2}}-\frac{2S_{-}S_{+}\left( q_{-}^{2}-q_{+}^{2}\right) ^{2}}{%
\delta q_{+}^{3}q_{-}^{3}}\right].
\end{equation*}%
It is seen that the critical current can be represented as the sum of two
terms. The first is the contributions from individual domains separated by
fully opaque FF wall %
\begin{equation}
\frac{eI_{C1}R_{N}}{2\pi T_{C}}=\frac{T}{T_{C}}\sum_{\omega >0}\frac{%
G_{0}^{2}\Delta ^{2}}{\gamma _{BS}\omega ^{2}}\Real\frac{1}{\sqrt{\widetilde{%
\Omega }_{+}}\sinh \left( d_{F}\sqrt{\widetilde{\Omega }_{+}}\right) },
\label{Ic1}
\end{equation}%
while the second%
\begin{equation}
\frac{eI_{C2}R_{N}}{2\pi T_{C}}=\frac{4h^{2}T}{Wd_{F}T_{C}}\sum_{\omega >0}%
\frac{G_{0}^{2}\Delta ^{2}}{\gamma _{BS}\omega ^{2}}\sum_{n=-\infty
}^{\infty }\frac{(-1)^{n}S_{-}S_{+}}{q_{+}^{3}q_{-}^{3}\delta }
\label{IcD}
\end{equation}%
gives the contribution from the domain wall. Here $\Real(a)$ denotes the
real part of $a.$\

Expression (\ref{Ic1}) reproduces the well-known result previously obtained
for single-domain SIFS structures \cite{Baladie}-\cite{Vasenko}
thereby demonstrating the independence of
the critical current on the orientation of the domains magnetization
vectors, if they are collinear oriented and the FF interface is fully
opaque for electrons.

\textbf{Limit of large  $\gamma _{BF}.$}
For large values of suppression parameter $\gamma _{BF}\gg \max \left\{
1,(Wq_{\pm })^{-1}\right\} $ expression (\ref{IcD}) transforms to%
\begin{equation}
\frac{eI_{C2}R_{N}}{2\pi T_{C}}=\frac{4h^{2}T}{Wd_{F}T_{C}}\sum_{\omega >0}%
\frac{G_{0}^{2}\Delta ^{2}}{\gamma _{BF}\gamma _{BS}\omega ^{2}}%
\sum_{n=-\infty }^{\infty }\frac{(-1)^{n}}{q_{+}^{4}q_{-}^{4}}.  \label{IcDL}
\end{equation}%
The sum over $n$ in Eq. (\ref{IcDL}) can be calculated analytically using
the theory of residues%
\begin{equation}
\frac{eI_{C2}R_{N}}{2\pi T_{C}}=\frac{2hT}{WT_{C}}\sum_{\omega >0}\frac{%
G_{0}^{2}\Delta ^{2}}{\gamma _{BF}\gamma _{BS}\omega ^{2}}S_{1},
\label{IcDLA}
\end{equation}%
\begin{equation*}
S_{1}=\Real\left[ \frac{i}{\widetilde{\Omega }_{+}^{3/2}}\left(
\frac{1}{\cosh \left( d_{F}\sqrt{\widetilde{\Omega }_{+}}\right) }+\frac{%
d_{F}\sqrt{\widetilde{\Omega }_{+}}}{\sinh \left( d_{F}\sqrt{\widetilde{%
\Omega }_{+}}\right) }\right) \right]
\end{equation*}%

\ It is seen that $I_{C2}$ is vanished as $(\gamma _{BF}W)^{-1}$ with
increase of $\gamma _{BF}W$ product and scales on the same characteristic lengths $\xi_{1},$
$\xi_{2}$ as the critical current  for single-domain SIFS structures (\ref{Ic1}).

\textbf{Limit of small  $\gamma _{BF}.$ }
In the opposite limit, $\gamma _{BF}\ll \max \left\{ 1,(Wq_{\pm
})^{-1}\right\} $ we have%
\begin{equation}
\frac{eI_{C2}R_{N}}{2\pi T_{C}}=\frac{8h^{2}T}{Wd_{F}T_{C}}\sum_{\omega >0}%
\frac{G_{0}^{2}\Delta ^{2}}{\gamma _{BS}\omega ^{2}}S_{2},  \label{IcSmG}
\end{equation}%
\begin{equation*}
S_{2}=\sum_{n=-\infty }^{\infty }\frac{(-1)^{n}S_{-}S_{+}}{%
q_{+}^{3}q_{-}^{3}\left( q_{-}C_{+}S_{-}+q_{+}S_{+}C_{-}\right) }.
\end{equation*}%

It is seen that in full agreement with the result obtained in \cite{Buzdin} in the
considered limit of large domain width, $W\gg \Real(q_{\pm }),$
\begin{equation}
\frac{eI_{C2}R_{N}}{2\pi T_{C}}=\frac{4h^{2}T}{Wd_{F}T_{C}}\sum_{\omega >0}%
\frac{G_{0}^{2}\Delta ^{2}}{\gamma _{BS}\omega ^{2}}\sum_{n=-\infty
}^{\infty }\frac{(-1)^{n}}{q_{+}^{3}q_{-}^{3}\left( q_{-}+q_{+}\right) }.
\label{Ic2LW}
\end{equation}%
contribution to the critical current from domain wall region falls as $%
W^{-1} $ and decays in the scale of $\xi _{1}.$

\textbf{Limit of small domain width.
} In the opposite case, $W\ll \Real(q_{\pm }),$ presentation of the critical
current as a sum of $I_{C1}$ and $I_{C2}$ is not physically reasonable and
for $I_{C}$ from (\ref{CurrF}) we get\
\begin{equation}
\frac{eI_{C}R_{N}}{2\pi T_{C}}=\frac{T}{2T_{C}}\sum_{\omega >0}\frac{%
G_{0}^{2}\Delta ^{2}}{\gamma _{BS}d_{F}\omega ^{2}}S_{3},  \label{Ic2AG}
\end{equation}%
\begin{equation*}
S_{3}=\sum_{n=-\infty }^{\infty }(-1)^{n}\left[ \frac{\left(
q_{-}^{2}+q_{+}^{2}\right) \gamma _{BW}+4}{\left( q_{-}^{2}q_{+}^{2}\gamma
_{BW}+q_{-}^{2}+q_{+}^{2}\right) }\right] ,
\end{equation*}%
where $\gamma _{BW}=\gamma _{BF}W/2.$ It is seen that for $\gamma _{BW}\gg 1$
expression (\ref{Ic2AG}) transforms to (\ref{Ic1}) and $I_{C}=I_{C1},$ while
in the limit $\gamma _{BW}\rightarrow 0$ from (\ref{Ic2AG}) it follows that the critical current
\begin{equation}
\frac{eI_{C}R_{N}}{2\pi T_{C}}=\frac{T}{T_{C}}\sum_{\omega >0}\frac{%
G_{0}^{2}\Delta ^{2}}{\gamma _{BS}\omega ^{2}\sqrt{\Omega }\sinh \left( d_{F}%
\sqrt{\Omega }\right) }  \label{SINS}
\end{equation}%
 is independent on exchange energy and falls with increase of $d_{F}$ in the
same scale as it is for SINS devices. Previously it was found that such transformation of
decay length takes place in a vicinity of domain wall
\cite{Chtchelkatchev} - \cite{Crouzy}. In particular, it was shown that if a sharp domain
wall is parallel \cite{Maleki}, \cite{Volkov2008} or perpendicular to SF
interface \cite{Crouzy} and
the thickness of ferromagnetic layers, $d_{f}$ $%
\lesssim \xi _{F},$ then for antiparallel direction of magnetization
the
 exchange field effectively averages out, and the decay length of
superconducting correlations becomes close to that of a
single nonmagnetic N
metal $\xi _{F}=\sqrt{D_{F}/2\pi T_{C}}.$
The same effect may also take place in S-FNF-S variable thickness bridges \cite{Karminskaya2},
\cite{Karminskaya4}.

For arbitrary values of $\gamma _{BW}$ the sum over $n$ in (\ref{Ic2AG}) can
be also calculated analytically. The denominator in (\ref{Ic2AG})  has the
poles at
\begin{equation*}
n=\pm i\frac{d_{F}}{\pi }\sqrt{\Omega +\frac{1\pm \sqrt{1-\gamma
_{BW}^{2}h^{2}}}{\gamma _{BW}}.}
\end{equation*}%
Application of the residue theorem to the summation of the series in $n$ in
the expression (\ref{Ic2AG}) leads to%
\begin{equation}
\frac{eI_{C}R_{N}}{2\pi T_{C}}=\frac{T}{2T_{C}}\sum_{\omega >0}\frac{%
G_{0}^{2}\Delta ^{2}}{\gamma _{BS}\omega ^{2}}\frac{\gamma _{BM}}{\sqrt{%
1-\gamma _{BM}^{2}h^{2}}}S_{4},  \label{IcAW}
\end{equation}%
\begin{equation*}
S_{4}=\frac{q}{\sqrt{\Omega +p}\sinh \left( d_{F}\sqrt{\Omega +p}\right) }-%
\frac{p}{\sqrt{\Omega +q}\sinh \left( d_{F}\sqrt{\Omega +q}\right) },
\end{equation*}%
\begin{equation}
p=\frac{1-\sqrt{1-\gamma _{BW}^{2}h^{2}}}{\gamma _{BW}},~q=\frac{1+\sqrt{%
1-\gamma _{BW}^{2}h^{2}}}{\gamma _{BW}}.  \label{pq}
\end{equation}%
It is seen that for\ $\gamma _{BW}h\leq 1$ s-wave superconducting
correlations 
decay exponentially into the F metal without any oscillations with two characteristic scales,  $\xi _{11}=\xi_{F}(\Omega+p)^{-1/2},$ and, $\xi _{12}=\xi_{F}(\Omega+q)^{-1/2}.$
If $\gamma _{BW}$ tends to zero then one of the damping characteristic
scale $\xi _{11}$ goes to that $\xi _{F}\Omega ^{-1/2}$ of SINF
junctions (see (\ref{SINS})), while the other $\xi _{12}$ goes to zero. With $\gamma _{BW}$ increase $%
\xi _{11}$  reduces, whereas $\xi _{12}$ increases, so that at $\gamma
_{BW}h=1$ they become equal to each other $\xi _{11}=\xi _{12}=\xi
_{F}\left( \Omega +h\right) ^{-1/2}.$ Further increase of $\gamma _{BW}h$
leads to appearance of the damped oscillations in $I_{C}(d_{F})$ dependence
with the ratio
\begin{equation}
\frac{\xi _{1}}{\xi _{2}}=\frac{\sqrt{\gamma _{BW}^{2}h^{2}-1}}{\sqrt{\left(
\gamma _{BW}\Omega +1\right) ^{2}+\gamma _{BW}^{2}h^{2}-1}+\Omega \gamma
_{BW}+1},  \label{ratio}
\end{equation}%
which monotonically increase from zero at $\gamma _{BW}h=1$ up to that of
single domain SIFS\ junctions
\begin{equation}
\frac{\xi _{1}}{\xi _{2}}=\frac{h}{\left( \sqrt{\Omega ^{2}+h^{2}}+\Omega
\right) },  \label{rdirty}
\end{equation}%
in the limit $\gamma _{BW}\rightarrow \infty .$

From (\ref{ratio}) (\ref{rdirty}) we can conclude that the existence of
domain structure in the F layer of SIFS\ devices can significantly modify
the relation between $\xi _{1}$ and $\xi _{2}$ extracted from experimental
studies of $I_{C}(d_{F})$ dependence in SIFS\ tunnel junctions.

This conclusion is valid not only in the limit of small domain
width.

\textbf{Arbitrary values of the domain width.}
For arbitrary values of the width of the magnetic domains to calculate the dependence of $I_{C}(d_{F})$ is necessary to use the general expression (\ref{CurrF}).  Figure 2 gives the $I_{C}(d_{F})$ curves calculated for $H=10 \pi T_{C},$ $\gamma_{BF}=0$ and
for a set of widths $W/ \xi_{F}.$ It is seen that in full accordance with the analytical analysis given above for $W$ smaller than $0.78\xi_{F},$ $I_{C}$ falls monotonically with $W$ increase. At $W\gtrsim0.78$ there is a transformation from a monotonic dependence of $I_{C}(d_{F})$ to a damped oscillatory one. It is interesting to note that in the vicinity of the transition the critical current decays even faster than for large $W.$
\begin{figure}[h]
\centerline{\includegraphics[width=0.5\textwidth]{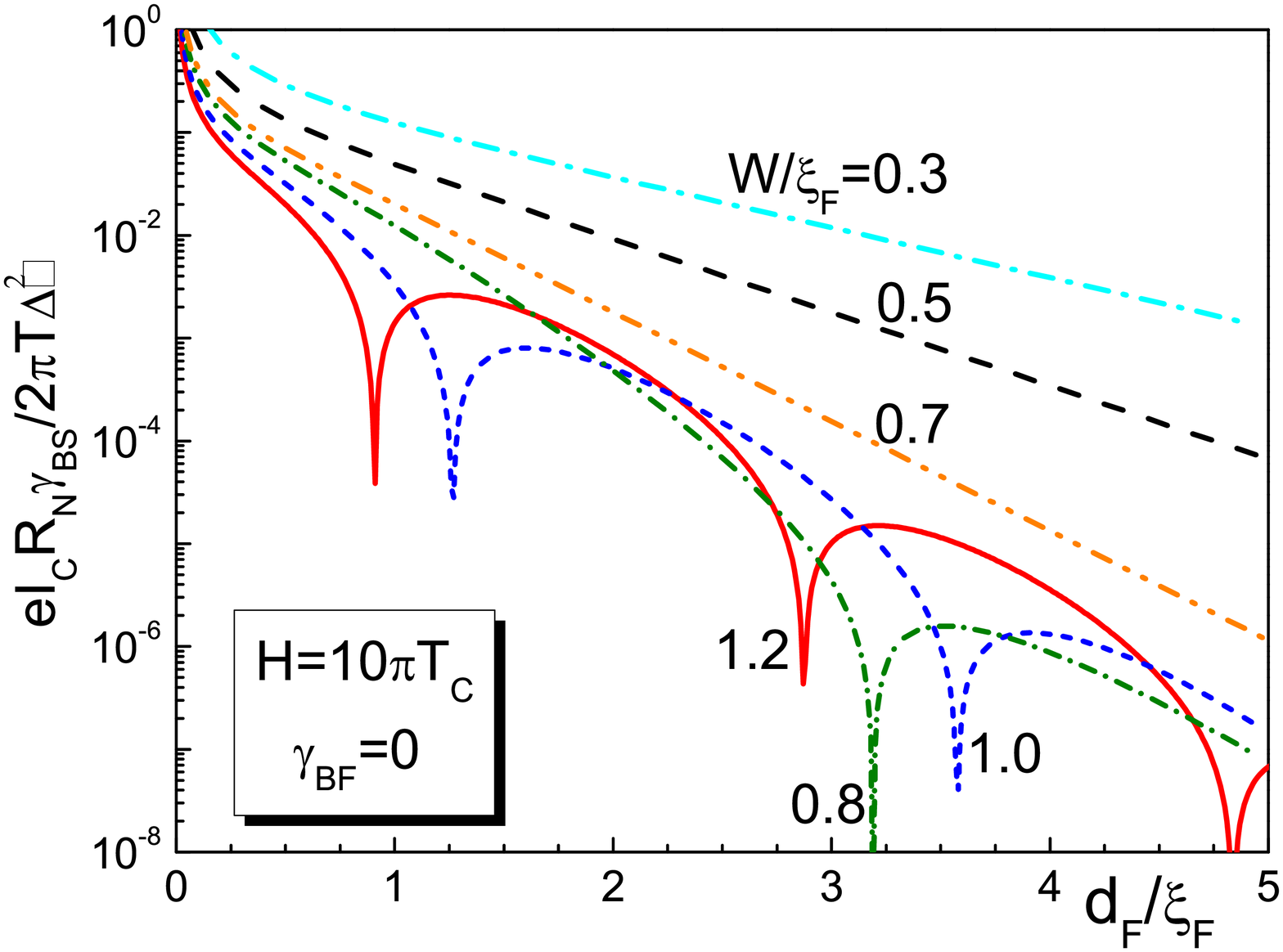}}
\caption{Fig.\protect\ref{fig:fig2}. Dependence of the critical current of SIFS Josephson junction as
a function of thickness of F layer $d_{F}$ calculated numerically from (\ref{CurrF}) for $T=0.5T_{C},$ $H=10 \pi T_{C},$ $\gamma_{BF}=0$ and
for a set of widths $W/ \xi_{F}=0,3; 0,5; 0,7; 0,8; 1; 1,2.$  }
\label{fig:fig2}
\end{figure}
To illustrate this result, we make a fit of the calculated curves by the simple expression
\begin{equation*}
I_{C}(d_{F})=A\exp (-d_{F}/\xi _{1})\cos (d_{F}/\xi _{2}+\varphi ),
\end{equation*}
which is ordinary used for estimation of the decay lengths $\xi_{1}$ and $\xi_{2}$ from an experimental data \cite{BK}, \cite{BR}.
At the first step we define $\xi _{2}\,$
\begin{equation*}
\xi _{2}=(d_{F2}-d_{F1})/\pi
\end{equation*}
from the positions of the first, $d_{F1},$ and the second, $d_{F2},$ $0$-$\pi $
transitions in $I_{C}(d_{F})$ dependence and put
\begin{equation*}
\varphi =\pi /2-d_{F1}/\xi _{2}
\end{equation*}
in order to get $I_{C}(d_{F1})=0.$
The decay length $\xi_{1}$ is determined from the ratio of magnitudes of
critical current  taken in two points having equal phase of oscillation:%
\begin{equation*}
\xi _{1}=\pi \xi _{2}\ln \left[ \frac{I_{C}(d_{F1}+\xi _{2}\pi /2)}{%
I_{C}(d_{F2}+\xi _{2}\pi /2)}\right]
\end{equation*}
and
normalization constant $A$
\begin{equation*}
A=\frac{I_{C}(d_{F1}+\xi _{2}\pi /2)}{\exp (-d_{F}/\xi _{1})\cos (d_{F}/\xi
_{2}+\varphi )}
\end{equation*}
has been determined by direct calculation of
magnitude in the certain point between $0$-$\pi $ transitions.
If the position of the second $0$-$\pi $ transition exceeds $10~\xi _{F}$, we
suppose that $\xi _{2}$ is infinite and $I_{C}(d_{F1})$ dependence can be fitted by function
\begin{equation*}
J_{C}(dF)=A\exp (-d_{F}/\xi _{1}).
\end{equation*}%
The results of the fitting procedure are presented in Fig.3-Fig.5, which give the decay lengths $\xi_{1}$ and $\xi_{2}$
as well as their ratio $\xi_{1}/ \xi_{2}$  calculated at $T=0.5T_{C},$ $H=10 \pi T_{C}$ for a set of suppression parameter
$\gamma_{BF}=0; 0.3; 1.$ Thin vertical lines in Fig.3, Fig.4 give values on the x-axis, at which there is a transition from a monotonous exponential decay of $I_{C}(d_{F})$ to the damped oscillation lows. Thin horizontal lines in Fig.3 - Fig.5 provide the asymptotic values of $\xi_{1},$ $\xi_{2}$
and $\xi_{1}/\xi_{2}$ in the limit $W \gg \xi_{F},$ which are coincide with the magnitudes calculated for single domain SIFS junction for given temperature $T=0.5T_{C}$ and exchange energy  $H=10\pi T_{C}.$

\begin{figure}[h]
\centerline{\includegraphics[width=0.5\textwidth]{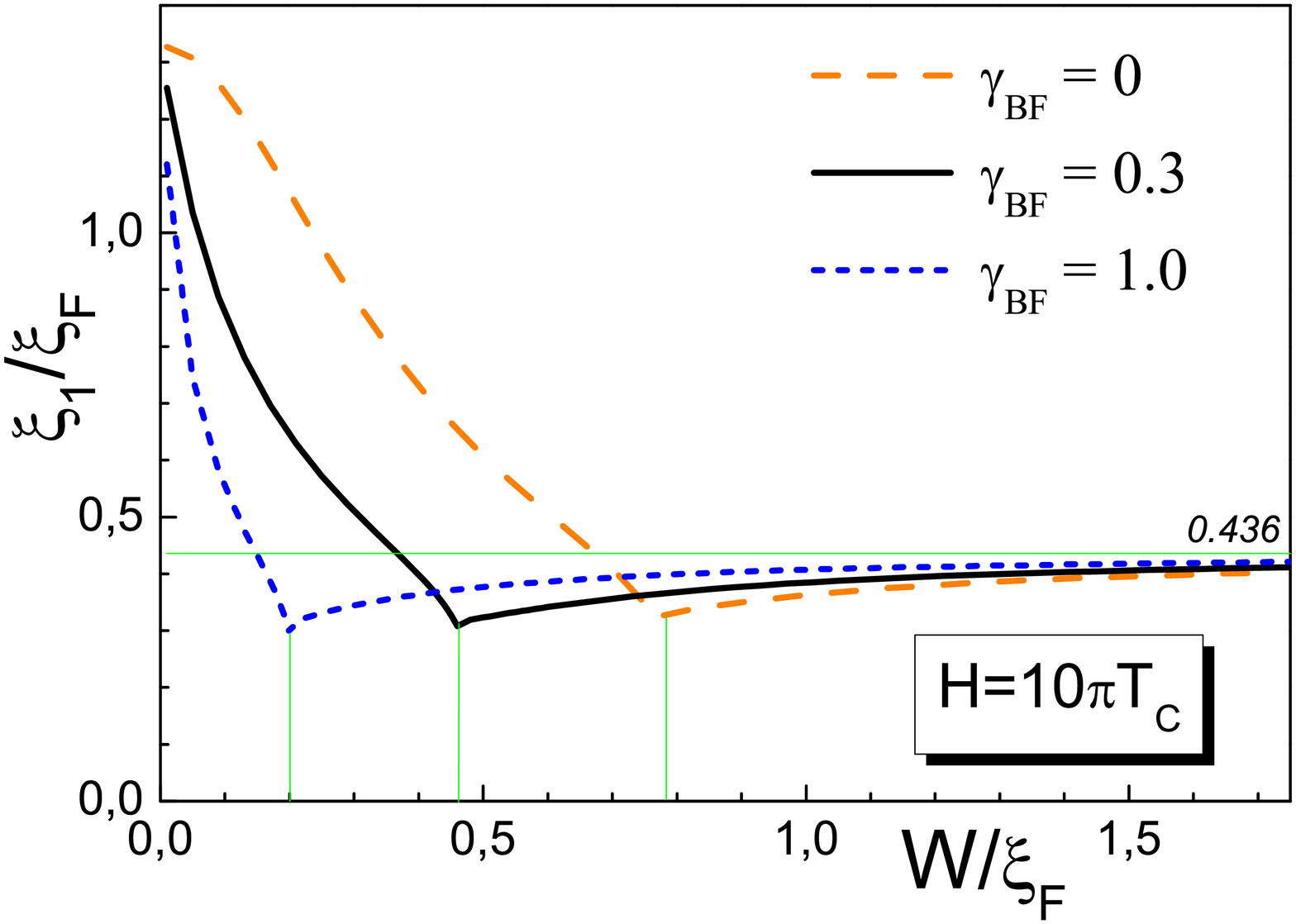}}
\caption{Fig.\protect\ref{fig:fig3}. Dependence of decay length $\xi_{1}$ as a function of domain width $W$ calculated at $T=0.5T_{C},$ $H=10 \pi T_{C}$ and
$\gamma_{BF}=0; 0.3; 1.$   }
\label{fig:fig3}
\end{figure}
\begin{figure}[h]
\centerline{\includegraphics[width=0.5\textwidth]{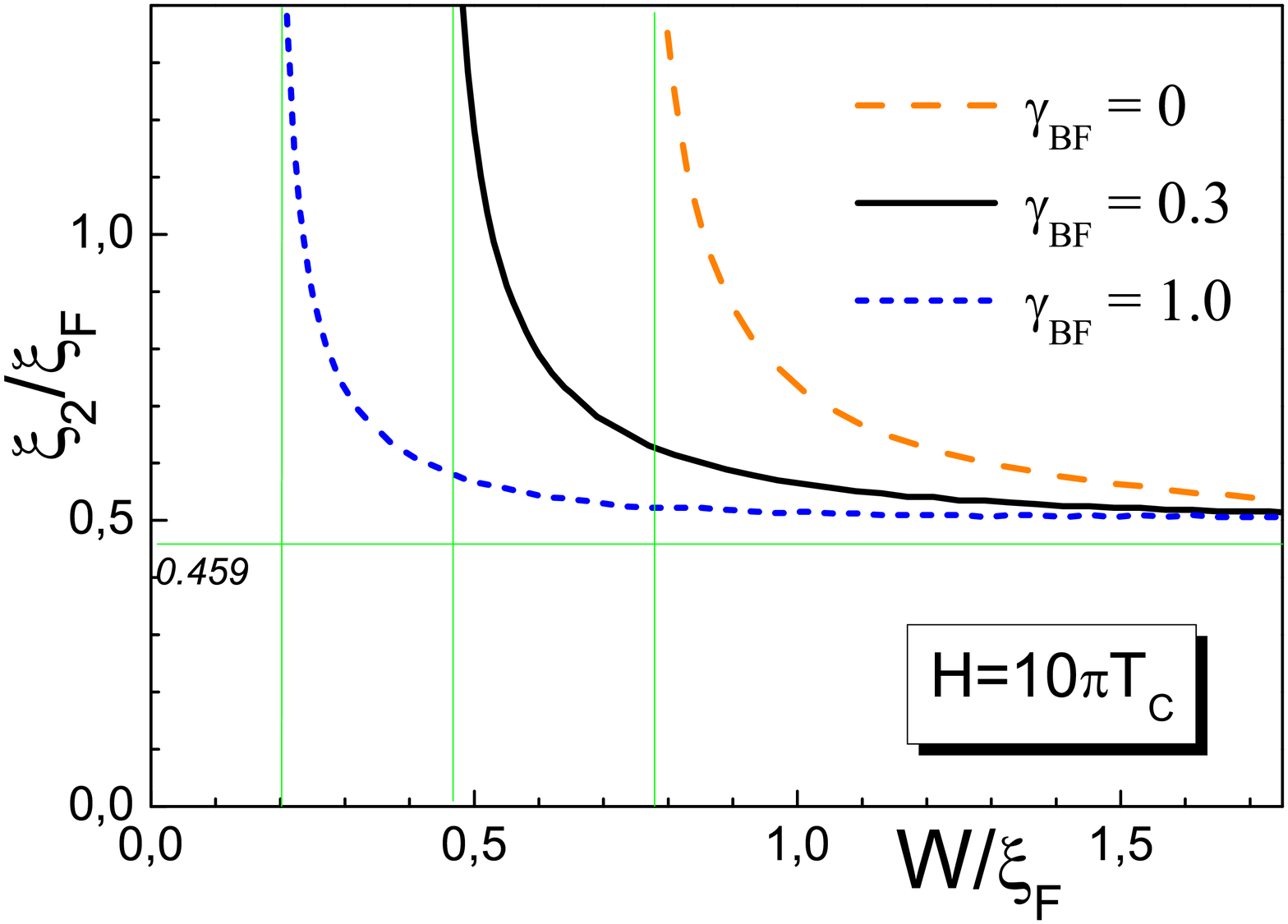}}
\caption{Fig.\protect\ref{fig:fig4}. Dependence of decay length $\xi_{2}$ as a function of domain width $W$ calculated at $T=0.5T_{C},$ $H=10 \pi T_{C}$ and
$\gamma_{BF}=0; 0.3; 1.$     }
\label{fig:fig4}
\end{figure}

It is seen that the transition point at which monotonic decay of $I_{C}(d_{F1})$ dependence transforms to a damped oscillation behavior
the smaller the larger is suppression parameter $\gamma_{BF}.$ Interestingly, in the vicinity of this transition decay length $\xi_{1}$ is even smaller compare to its magnitude in the limit of large $W.$
\begin{figure}[h]
\centerline{\includegraphics[width=0.5\textwidth]{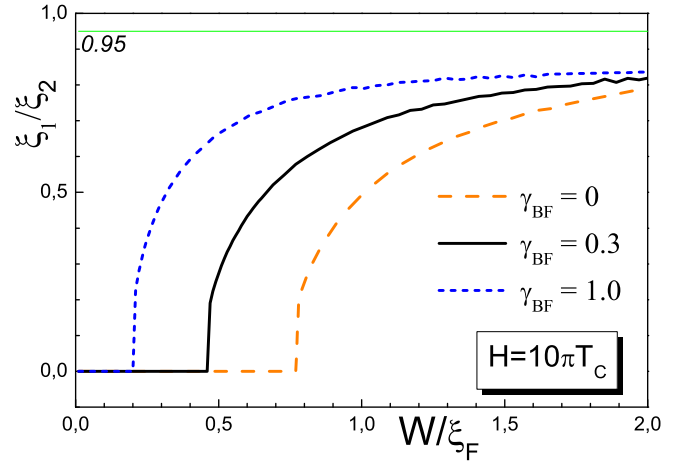}}
\caption{Fig.\protect\ref{fig:fig5}. The ratio  of decay lengths $\xi_{1}$ and $\xi_{2}$ as a function of domain width $W$ calculated at $T=0.5T_{C},$  $H=10 \pi T_{C}$ and
$\gamma_{BF}=0; 0.3; 1.$    }
\label{fig:fig5}
\end{figure}

It is also necessary to note that despite of the fact that the transition takes place at $W<\xi_{F},$ the difference between $\xi_{1}$ and $\xi_{2},$
as it follows from Fig.5, exists even for large domain width: the ratio $\xi_{1}/\xi_{2}$ is only around $0.8$ at $W=4\xi_{F}$ and very slowly tends to the following from (\ref{rdirty}) the single domain value $0.95$ with $W$ increase. This fact permits us to conclude that the difference between $\xi_{1}$ and $\xi_{2}$ experimentally observed in SFS Josephson structures based on dilute magnetic alloys can be also the consequence of existence of magnetic domains in the F layer.

This work was supported by RFBR grants l4-02-90018-bel$\_$a, 14-02-31002-mol$%
\_$a, 15-32-20362-mol$\_$a$\_$ved, Ministry of Education and Science of the Russian Federation in
the frameworks of Grant No. 14.587.21.0006 (RFMEFI58714X0006) and the Program for the Promotion of
Competitiveness of the Kazan Federal University
among the World-Leading Scientific Educational
Centers, Russian President grant MK-1841.2014.2, Dynasty Foundation,
Scholarship of the President of the Russian Federation and Dutch FOM. A.A. Golubov is also acknowledge EU COST program MP1201.

\vfill\eject

Fig.1. Geometry of the considered SIFS Josephson junction
and its enlarged part, which includes  two halves of domains and domain wall separating them.
The  insulating barrier I has a small transparency  (shown by a blue
line).

Fig.2. Dependence of the critical current of SIFS Josephson junction as
a function of thickness of F layer $d_{F}$ calculated numerically from (\ref{CurrF}) for $T=0.5T_{C},$ $H=10 \pi T_{C},$ $\gamma_{BF}=0$ and
for a set of widths $W/ \xi_{F}=0,3; 0,5; 0,7; 0,8; 1; 1,2.$

Fig.3. Dependence of decay length $\xi_{1}$ as a function of domain width $W$ calculated at $T=0.5T_{C},$ $H=10 \pi T_{C}$ and
$\gamma_{BF}=0; 0.3; 1.$

Fig.4. Dependence of decay length $\xi_{2}$ as a function of domain width $W$ calculated at $T=0.5T_{C},$ $H=10 \pi T_{C}$ and
$\gamma_{BF}=0; 0.3; 1.$

Fig.5. The ratio  of decay lengths $\xi_{1}$ and $\xi_{2}$ as a function of domain width $W$ calculated at $T=0.5T_{C},$  $H=10 \pi T_{C}$ and
$\gamma_{BF}=0; 0.3; 1.$


\bigskip
\begin{thebibliography}{99}
\bibitem{RevG} A.~A.~Golubov, M.~Yu.~Kupriyanov, E.~Il'ichev, Rev. Mod.
Phys. \textbf{76}, 411 (2004).

\bibitem{RevB} A.~I.~Buzdin, Rev. Mod. Phys. \textbf{77}, 935 (2005).

\bibitem{RevV} F.~S.~Bergeret, A.~F.~Volkov, K.~B.~Efetov, Rev. Mod. Phys.
\textbf{77}, 1321 (2005).

\bibitem{Kontos} T. Kontos, M. Aprili, J. Lesueur, F. Genet, B. Stephanidis,
and R. Boursier, Phys. Rev. Lett. \textbf{89}, 137007 (2002).

\bibitem{Bell} C. Bell, R. Loloee, G. Burnell, and M. G. Blamire, Phys. Rev.
B \textbf{71}, 180501 (R) (2005).

\bibitem{Shelukhin} V. Shelukhin, A. Tsukernik, M. Karpovski, \textit{et al.}%
, Phys. Rev. B \textbf{73}, 174506 (2006).

\bibitem{ryazanov1} V. A. Oboznov, V. V. Bol'ginov, A. K. Feofanov, V. V.
Ryazanov, and A. Buzdin, Phys. Rev. Lett. \textbf{96}, 197003 (2006).

\bibitem{Blamire} J. W. A. Robinson, S. Piano, G. Burnell, C. Bell, and M.
G. Blamire, Phys. Rev. Lett. \textbf{97}, 177003 (2006).

\bibitem{Bannykh} A. A. Bannykh, J. Pfeiffer, V. S. Stolyarov, I. E. Batov,
V. V. Ryazanov, and M. Weides, Phys. Rev. B \textbf{79}, 054501 (2009).

\bibitem{Born} F. Born, M. Siegel, E.K. Hollmann, H. Braak, A.A. Golubov,
D.Yu Gusakova, and M.Yu Kupriyanov, Phys. Rev. B \textbf{74}, 140501 (2006).

\bibitem{Robinson} J. W. A. Robinson, F. Chiodi, M. Egilmez, G. B. Halasz,
M. G. Blamire, Sci. Rep. \textbf{2}, 00699 (2012)

\bibitem{Blum} Y. Blum, A. Tsukernik, M. Karpovski, and A. Palevski, Phys.
Rev. B \textbf{70}, 214501 (2004).

\bibitem{Pugach} N.G. Pugach, M.Yu Kupriyanov, E. Goldobin, R. Kleiner, and
D. Koelle, Phys. Rev. B \textbf{84}, 144513 (2011).

\bibitem{Usadel} K.~D.~Usadel, Phys.\ Rev.\ Lett.\ \textbf{25}, 507 (1970).

\bibitem{KL} M.\,Yu.\ Kuprianov and V.\,F.\ Lukichev, Zh.\ Eksp.\ Teor.\
Fiz.\ \textbf{94}, 139 (1988) [Sov.\ Phys.\ JETP \textbf{67}, 1163 (1988)].

\bibitem{Baladie} A. Buzdin and I. Baladie, Phys. Rev. B \textbf{67}, 184519
(2003).

\bibitem{Faure} M. Faure, A. I. Buzdin, A. A. Golubov, and M. Yu.
Kupriyanov, Phys. Rev. B 73, 064505 (2006).

\bibitem{Vasenko} A.S. Vasenko, A.A. Golubov, M.Yu Kupriyanov, and M.
Weides, Phys. Rev. B \textbf{77}, 134507 (2008).

\bibitem{Buzdin} A. I. Buzdin, A. S. Mel'nikov, and N. G. Pugach, Phys. Rev.
B \textbf{83}, 144515 (2011).

\bibitem{Chtchelkatchev} N. M. Chtchelkatchev and I. S. Burmistrov, Phys.
Rev. B \textbf{68}, 140501(R) (2003).

\bibitem{HouzetBuz} M. Houzet and A. I. Buzdin, Phys. Rev. B \textbf{74},
214507 (2006).

\bibitem{Maleki} M. A. Maleki and M. Zareyan, Physical Review B \textbf{74},
144512 (2006).

\bibitem{Burmistrov} I. S. Burmistrov and N. M. Chtchelkatchev, Phys. Rev. B
\textbf{72}, 144520 (2005).

\bibitem{Volkov2008} A.F. Volkov, K.B. Efetov, Phys Rev B \textbf{78},
024519 (2008).

\bibitem{Soloviev} I.I. Soloviev, N.V. Klenov, S.V. Bakursky, M.Yu
Kupriyanov, A.A. Golubov, Pis'ma Zh. Eksp. Teor. Fiz. \textbf{101}, 258
(2015) {[}JETP Lett.\textbf{101}, 240 (2015){]}.

\bibitem{sfi2} B. Crouzy, S. Tollis, D. A. Ivanov, Phys. Rev. B \textbf{75},
054503 (2007).

\bibitem{Linder} I. B. Sperstad, J. Linder, and A. Sudbo, Phys. Rev. B
\textbf{78}, 104509 (2008).

\bibitem{Linder2} J. Linder and K. Halterman, Phys. Rev. B \textbf{90},
104502 (2014).

\bibitem{Baker} T. Baker, A. Richie-Halford, and A. Bill, New J. Phys.
\textbf{16}, 093048 (2014).

\bibitem{bh8} Ya. M. Blanter and F. W. J. Hekking, Phys. Rev. B \textbf{69},
024525 (2004).

\bibitem{Champel1} T. Champel and M. Eschrig, PRB \textbf{72}, 054523 (2005).

\bibitem{Fominov1} Ya. V. Fominov, A. F. Volkov, and K. B. Efetov, Phys.
Rev. B  \textbf{75}, 104509 (2007).

\bibitem{Crouzy} B. Crouzy, S. Tollis, D. A. Ivanov, Phys. Rev. B \textbf{76}%
, 134502 (2007).

\bibitem{Karminskaya2} T. Yu. Karminskaya and M. Yu. Kupriyanov, Pis'ma Zh.
Eksp. Teor. Fiz. \textbf{85}, 343 (2007) {[}JETP Lett. \textbf{86}, 61 (2007)%
{]}.

\bibitem{Karminskaya4} T. Yu. Karminskaya, A. A. Golubov, M. Yu. Kupriyanov,
and A. S. Sidorenko, Phys. Rev. B \textbf{79}, 214509 (2009).

\bibitem{BK} A.~I.~Buzdin, and M. Yu. Kupriyanov,  Pis'ma Zh.
Eksp. Teor. Fiz. \textbf{53}, 308 (1991) {[}JETP Lett. \textbf{53}, 321 (1991)%
{]}.

\bibitem{BR} A.~I.~Buzdin, V. V. Ryazanov, Physica C \textbf{460}, 238 (2007).



\end{thebibliography}
\end{document}